\documentclass[prl,amsmath,amssymb,onecolumn,superscriptaddress]{revtex4-1}
\usepackage{graphicx,bm,color,xcolor,appendix}
\usepackage{ulem}
\usepackage[colorlinks,bookmarks=true,citecolor=blue,linkcolor=blue,urlcolor=blue]{hyperref}

\begin{document}
\title{All-optical control of the photonic Hall lattice in a pumped waveguide array}
\author{Shirong Lin}
\email{shironglin@sjtu.edu.cn}
\affiliation{State Key Laboratory of Advanced Optical Communication Systems and Networks, School of Physics and Astronomy, Shanghai Jiao Tong University, Shanghai 200240, China}
\author{Luojia Wang}
\affiliation{State Key Laboratory of Advanced Optical Communication Systems and Networks, School of Physics and Astronomy, Shanghai Jiao Tong University, Shanghai 200240, China}
\author{Luqi Yuan}
\email{yuanluqi@sjtu.edu.cn}
\affiliation{State Key Laboratory of Advanced Optical Communication Systems and Networks, School of Physics and Astronomy, Shanghai Jiao Tong University, Shanghai 200240, China}
\author{Xianfeng Chen}
\affiliation{State Key Laboratory of Advanced Optical Communication Systems and Networks, School of Physics and Astronomy, Shanghai Jiao Tong University, Shanghai 200240, China}
\affiliation{Shanghai Research Center for Quantum Sciences, Shanghai 201315, China}
\affiliation{Jinan Institute of Quantum Technology, Jinan 250101, China}
\affiliation{Collaborative Innovation Center of Light Manipulations and Applications, Shandong Normal University, Jinan 250358, China}

\begin{abstract}
Quantum Hall system possesses topologically protected edge states which have enormous theoretical and practical implications in both fermionic and bosonic systems. Harnessing the quantum Hall effect in optical platforms with lower dimensionality is highly desirable with synthetic dimensions and has attracted broad interests in the photonics society. Here, we introduce an alternative way to realize the artificial magnetic field in a frequency dimension, which is achieved in a pump-probe configuration with cross-phase modulations in a one-dimensional four waveguide array.
The dynamics of the topological chiral edge state has been studied and the influence from the crosstalk of the pump fields has been explored. Our work shows an all-optical way to simulate the quantum Hall system in a photonic system and holds potential applications in manipulating light in waveguide systems.
\end{abstract}

\maketitle

\section{Introduction}
As the well-known topological system – quantum Hall system has attracted great attentions since its discovery at 1980\cite{klitzing1980new}, and stimulates the flourishing field of topological insulators\cite{xiao2010berry,ren2016topological}.
Simulating quantum Hall effect in photonics has experienced rapid progress in the past decade, which significantly deepens people's physical understanding and also offers rich potential applications towards generations of optical devices\cite{sun2017two,khanikaev2017two,ozawa2019topological,wu2017applications,kremer2021topological}.  The photonic analog of quantum Hall system has been successfully demonstrated in such various platforms as  photonic crystals\cite{haldane2008possible,wang2008reflection}, coupled resonator optical waveguides\cite{hafezi2013imaging,mittal2016measurement}, polaritonic \cite{karzig2015topological} and optomechanical systems\cite{schmidt2015optomechanical}.
In particular, the realization of artificial magnetic field\cite{fang2012realizing,fang2012photonic,fang2013experimental,tzuang2014non} provides an alternative way to simulating topological photonics systems.

However, it is fundamentally challengeable to generate the effective magnetic field in real spaces, and, as alternative ways, there are several methods developed in the synthetic space\cite{luo2015quantum, yuan2016photonic,ozawa2016synthetic,lustig2019photonic,dutt2020single,leefmans2021topological}.
The concept of synthetic dimension has been widely studied in different optical systems\cite{yuan2018synthetic,ozawa2019topologicalsynthetic,lustig2021topological}. Among these, synthetic dimension based on the frequency axis of light has experienced an increasing amount of investigations\cite{bell2017spectral,qin2018spectrum,dutt2019experimental,li2021dynamic,chen2021real,wang2021generating,yuan2021tutorial}, and turns out to be one of the most promising candidates for emulating topological physics. The original theoretical proposal in ring resonators utilizes multiple resonant modes connected through dynamic modulation of the refractive index inside the ring\cite{yuan2016photonic,ozawa2016synthetic}. On the other hand, nonlinear all-optical techniques have been exploited to achieve one-dimensional frequency dimension either by cross-phase modulation\cite{peschel2008discreteness,bersch2009experimental,bersch2011spectral} or four-wave mixing procedure\cite{bell2017spectral}. These all-optically pumped systems present a fast and efficient approach to manipulating physics in a synthetic dimension. 
Nevertheless, in realizing the quantum Hall model in all-optically pumped systems with synthetic dimensions might be affected by the crosstalk from separated pump optical fields. Understanding this effect can thus shed new light on exploring topological phases in all-optically pumped systems.

Here, we propose a viable all-optical model in a one-dimensional waveguide array to create a photonic quantum Hall system in a synthetic space including the frequency axis of light. In particular, we consider the light propagation inside four coupled waveguides, each of which is driven by a pump laser so a synthetic two-dimensional lattice is configured through cross-phase modulations\cite{lin2007nonlinear,agrawal2007nonlinear}.
We then find an alternative implementation of the effective magnetic field based on independently designing the phases of pump laser fields. In this manner, we create a photonic analog of quantum Hall model. Moreover, we identify the parameters for the crosstalk of pump lasers supporting the chiral edge states and show phase transitions to parameter regions where the topology breaks down. Importantly, our scheme not only simulates the optical counterpart of a nontrivial topological system, but also shows an all-optical way to control the topological system, which could be useful in developing laser-based technologies\cite{chen2021highlighting}.

\section{Theoretical model}
It has been shown that a probe field propagating inside a single waveguide under the influence of a co-propagating pump can construct the synthetic frequency dimension via cross-phase modulation\cite{peschel2008discreteness}. We first summarize such behavior here. The evolution of the probe-field amplitude $u^s$ along the $z$ direction can be described by \cite{peschel2008discreteness,agrawal2007nonlinear}
\begin{equation}
  \left(i\dfrac{\partial}{\partial z}+i\Delta\dfrac{\partial}{\partial t}
  +V(t)\right)u^s=0,\label{refeq1}
\end{equation}
where $\Delta$ represents the group-velocity mismatch between the pump and the probe fields. $V(t)=|u^p (t)|^2$ describes the time-dependent effective potential for the probe field propagating through the waveguide. In deriving Eq. (\ref{refeq1}), a reference frame where the pump field is at rest has been chosen, which results in the walk-off mismatching parameter $\Delta$, and the probe field is assumed to be weak enough to neglect its self-phase modulation.

Although the shape of the pump field as the writing laser beam can take any form, of most important interest to us is the periodical cosine-shape form as $V(t)=P_0\left(\cos(\Omega t)+1\right)$ with $P_0$ being the peak power of the periodic field and $\Omega$ being the modulation frequency,
which can be utilized to construct a synthetic lattice structure in the frequency dimension\cite{peschel2008discreteness}. Therefore, Eq. (\ref{refeq1}) describes the dynamics of the probe field in a dimension along the frequency axis of light. This fact can been seen if one Fourier-transforms the Eq. (\ref{refeq1}) with respect to time as $U^s(\omega)=\int dtu^s (t)$exp$(i\omega t)/2\pi$ ,which yields:
\begin{equation}
\label{refeq2}
  \left(i\dfrac{\partial}{\partial z}+\omega\Delta
  +2\gamma P_0\right)U^s(\omega)+\gamma P_0\left[U^s(\omega-\Omega)+U^s(\omega+\Omega)\right] =0,
\end{equation}
where $\gamma$ denotes the nonlinear coefficient of cross-phase modulation. 
Furthermore, by the introduction of discrete spectral amplitudes
 $a_{n}=U^s(\omega_0+n\Omega, z)$exp$[-i(\omega_0\Delta+2\kappa)z]$,
we obtain\cite{peschel2008discreteness} 
\begin{equation}
  \left(i\dfrac{\partial}{\partial z}+n\Delta\Omega
  \right)a_{n}+\kappa\left(a_{n-1}+a_{n+1}\right) =0,\label{refeq3}
\end{equation}
where the modulation strength $\kappa\equiv\gamma P_0$.
Eq. (\ref{refeq3}) reveals that the spectral component of the probe field at a frequency $\omega_n \equiv \omega_0 + n\Omega$, where $\omega_0$ is a reference frequency and $n$ is an integer,
 couples to two nearby spectral components at $\omega_0+(n\pm 1)\Omega$.
As such, a synthetic frequency dimension for the probe propagating along $z$ is constructed utilizing the sideband generation induced by the cross-phase modulation from the pump field.

\begin{figure*}
\centering\includegraphics[width=8.2cm]{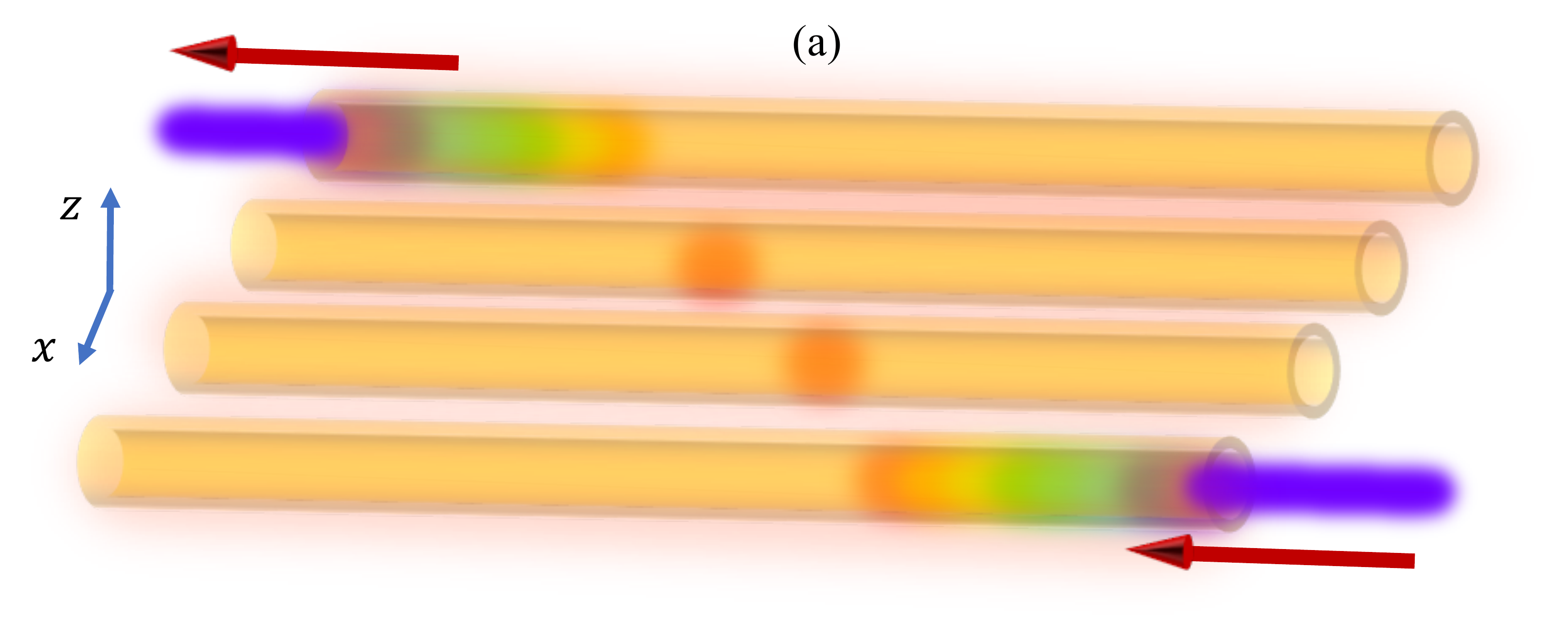}
\centering\includegraphics[width=6.0cm]{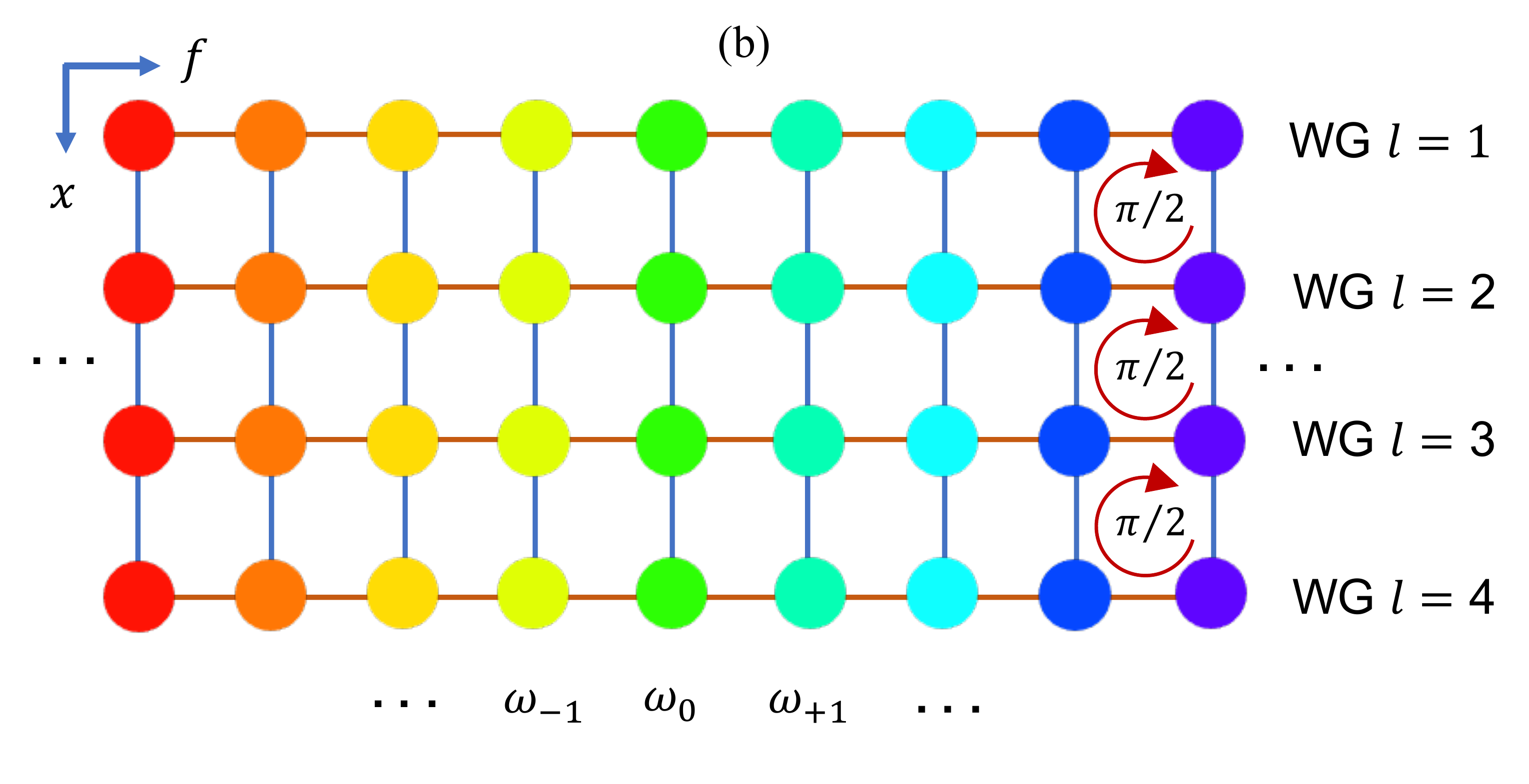}
\caption{(Color online) (a) The schematic of probes propagating in the one-dimensional pumped waveguide array. (b) The synthetic two-dimensional lattice consisting a spatial and a frequency dimensions, corresponding to the waveguide array in (a). WG denotes to waveguide.}
\label{fig1}
\end{figure*}

Based on the all-optical construction of a frequency dimension, we now show that a photonic Hall lattice in a pumped waveguide array can be realized. Light propagation in the synthetic dimensions is illustrated schematically in Fig. \ref{fig1}(a). Our system consists of four waveguides which form a one-dimensional array. A probe field propagates along the waveguide array under the influence of a co-propagating pump in each waveguide via cross-phase modulations. Under a tight-binding description, the probe field hops between nearest-neighbor waveguides. In such a nonlinear optical system the evolution of the probe field is described by 
\begin{equation}
  \left(i\dfrac{\partial}{\partial z}+i\Delta\dfrac{\partial}{\partial t}
  +V_l(t)\right)u^s_{l}+g\left(u^s_{l-1}+u^s_{l+1}\right)=0,\label{seq1}
\end{equation}
where $g$ is the coupling coefficient of the probe field when it is propagating in the pumped waveguide array, and $l=1,2,3,4$. 
Again, applying the same procedure as discussed above, we find that the dynamics of the probe field is given by
\begin{equation}
\label{seq2}
  \left(i\dfrac{\partial}{\partial z}+n\Delta\Omega
  \right)a_{n,l}+\kappa\left(a_{n-1,l}+a_{n+1,l}\right)+g\left(a_{n,l-1} +a_{n,l+1}\right)=0,
\end{equation}
In fact, Eq. (\ref{seq2}) describes the dynamics of the probe field in synthetic two-dimensional space which consists of a spatial dimension and a frequency dimension. Hence the equivalence of a two-dimensional lattice shows up.

The strategy to obtain an all-optical equivalent of quantum Hall system in the synthetic dimension is to introduce phase modulation for the individual pump fields. Specifically, for a scheme of $(0, \pi/4, \pi/2, 3\pi/4)$ assigned to the phase of the pump fields in waveguides 1-4 accordingly, we are led to introduce the phase structure of $(0, \pi/2, \pi, 3\pi/2)$ sensed by the probe according to the cross-phase modulations. 
In Eq.(\ref{seq1}), explicitly, we thus have $V_l(t)=P_0\left(\cos(\Omega t+\phi_l)+1\right)$ with $\phi_l=(0, \pi/2, \pi, 3\pi/2)$. Therefore, in a pumped four-waveguide system, the $\pi/2$ phase-difference between the neighboring waveguides implements the Landau gauge, and gives rise to a magnetic flux equal to $1/4$ per plaquette\cite{hofstadter1976energy}. In this way, the equation of motion of the probe field becomes 
 \begin{equation}
 \label{eom}
  \left(i\dfrac{\partial}{\partial z}+n\Delta\Omega
  \right)a_{n,l}+\kappa\left(a_{n-1,l}e^{i\phi_l}+a_{n+1,l}e^{-i\phi_l}\right)+g\left(a_{n,l-1}+a_{n,l+1}\right)=0,
\end{equation}
which corresponds to a charged particle on a square lattice subjected to a magnetic field and an electric field. This is analogous to the Schr\"odinger equation where the roles of time, $t$, are replaced by the propagation direction, $z$. If the effects resulting from the electric field is negligible, i.e., $n\Delta\Omega\ll\kappa$, we can focus only on the magnetic field, which gives rise to the corresponding Hamiltonian 
 \begin{equation}
 \label{seq4}
  H=\sum_{\substack{n,l}} \left[\kappa \left(e^{-i\phi_l}a_{n,l}^\dag a_{n+1,l}+e^{i\phi_l}a_{n+1,l}^\dag a_{n,l}\right)+g\left(a_{n,l}^\dag a_{n,l+1}+a_{n,l+1}^\dag a_{n,l}\right)\right],
\end{equation}
where $a_{n,l}^\dag (a_{n,l})$ is the bosonic creation(annihilation) operator. This Hamiltonian informs that the frequency space contributes to one of the dimension (labeled by the $n$), as shown in Fig. \ref{fig1}(b). 
We emphasize that, under the all-optical configuration, the crosstalk from separated pump optical fields in nearby waveguides come into play and have a role in the dynamics of the edge states, which therefore requires careful examinations. Following the coupled mode theory\cite{yariv1973coupled,haus1991coupled,huang1994coupled}, we can write the evolution of pump fields along each waveguides with nearest-neighbor couplings 
\begin{equation}
i\dfrac{\partial}{\partial z}
  u^p_{l}+g'\left(u^p_{l-1}+u^p_{l+1}\right)=0,\label{peq1}
\end{equation}
where $g'$ represents the coupling coefficient for the pump fields. Consequently, we obtain position-dependent pumps power $P_l(z)=|u^p_{l}(z)|^2$, and thus have a position-dependent modulation strength $\kappa_l(z)\equiv\gamma P_l(z)$, together with the phases $\phi_l(z)$ changing accordingly for the probe field propagating in the $l$-th waveguide. Therefore, the working Hamiltonian dependent on propagation distance $z$ becomes
\begin{equation}
\label{seq6}
  H(z)=\sum_{\substack{n,l}} \left[\kappa_l(z) \left(e^{-i\phi_l(z)}a_{n,l}^\dag a_{n+1,l}+e^{i\phi_l(z)}a_{n+1,l}^\dag a_{n,l}\right)+g\left(a_{n,l}^\dag a_{n,l+1}+a_{n,l+1}^\dag a_{n,l}\right)\right].
\end{equation}
In view of the inevitable crosstalk of pump fields, we point out that understanding the role of the influence of pump fields in such all-optical system not only is necessary, but also could provide a flexible way to manipulate light in such photonic Hall lattice.

\section{Results and discussion}
\subsection{Simulation of quantum Hall effects in the synthetic space}\label{sec3.1}
 \begin{figure*}
\centering\includegraphics[width=6.0cm]{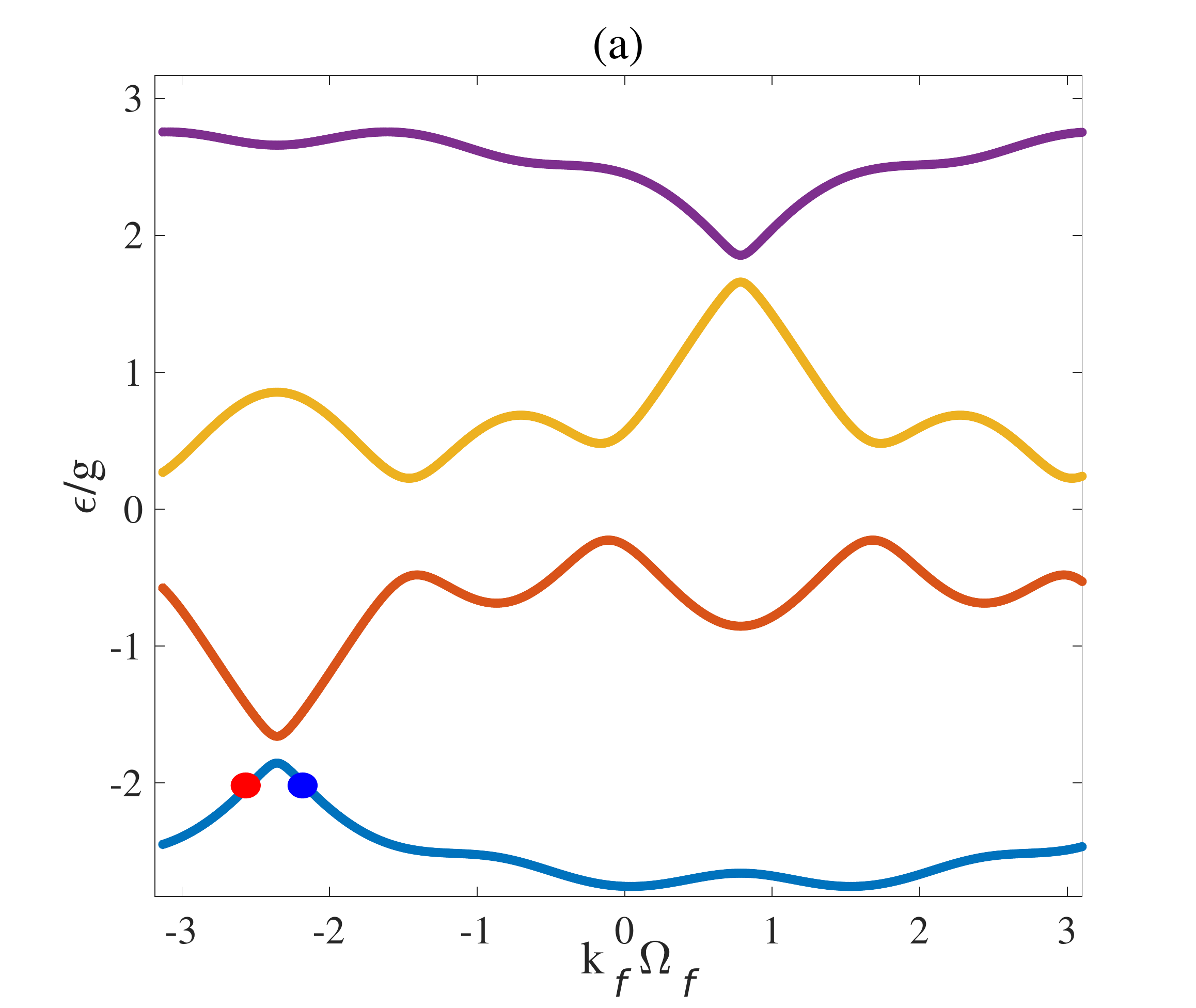}
\centering\includegraphics[width=6.0cm]{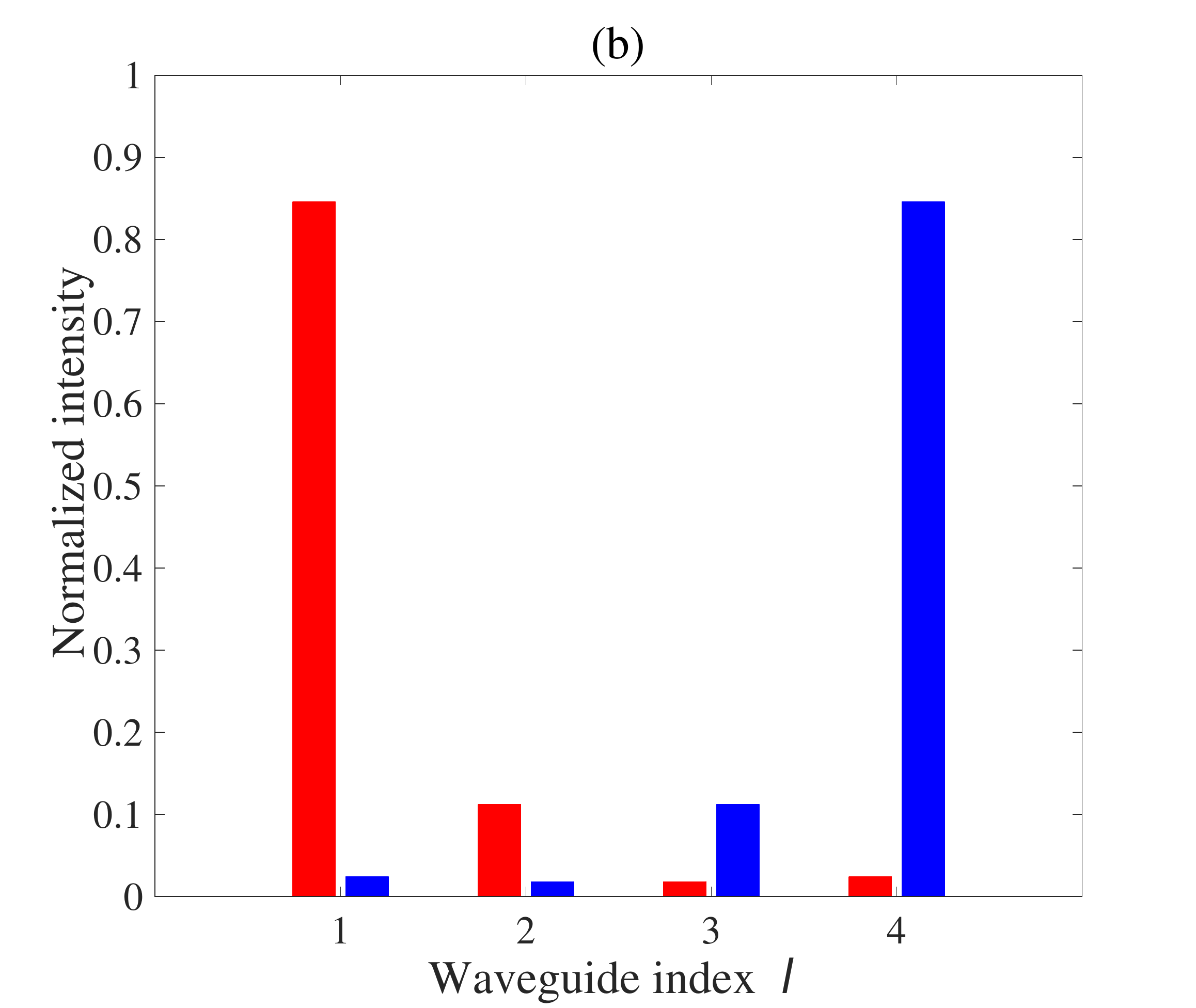}
\centering\includegraphics[width=6.0cm]{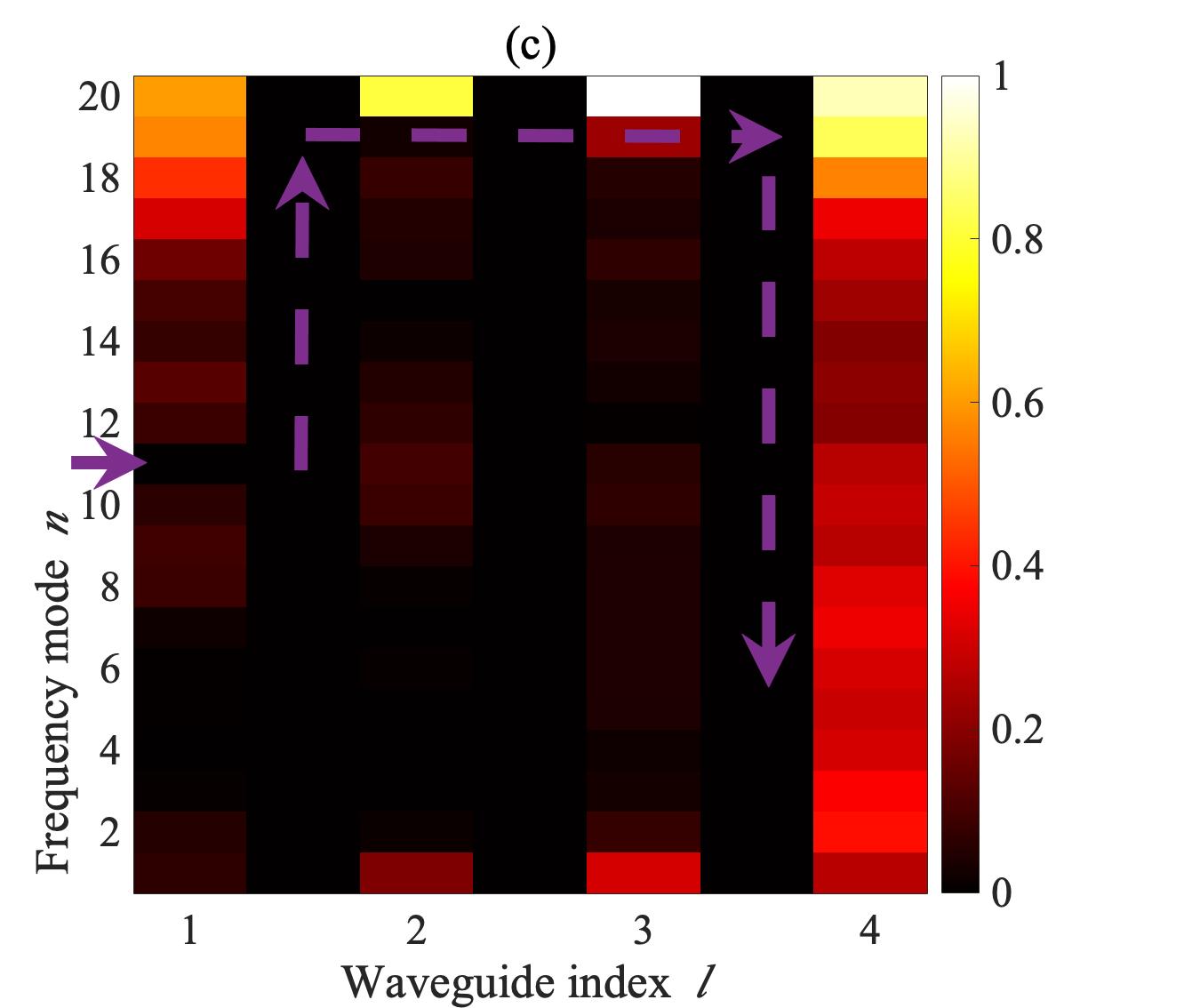}
\centering\includegraphics[width=6.0cm]{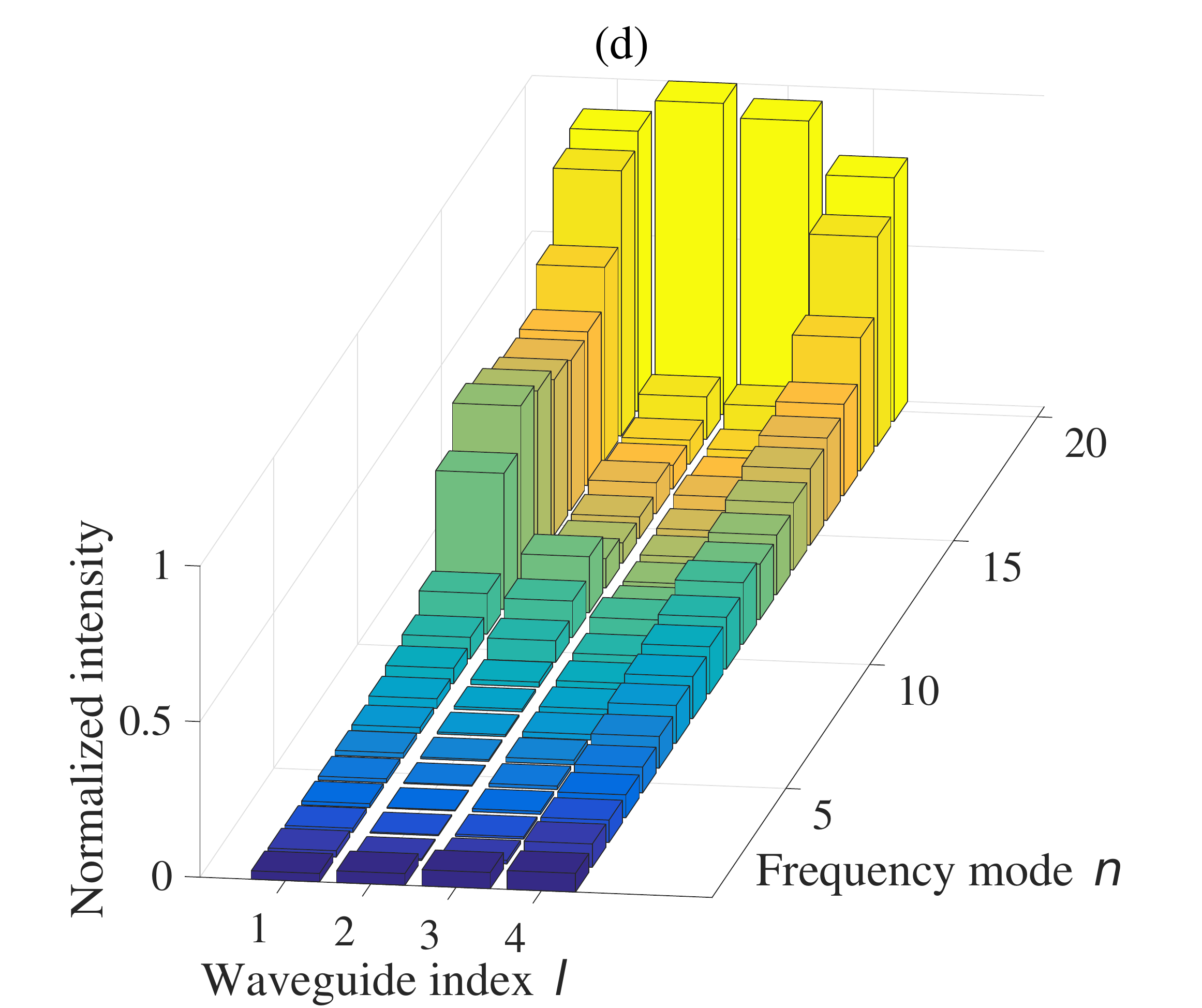}
\caption{(Color online) (a) Projected band structure  versus $k_f$ for the four-waveguide system. (b) Field intensity distributions of the two edge states indicated by red and blue dots in (a), which are localized on the two boundaries and decay exponentially into the bulk. (c) Distributuion of $|v_{n,l}|^2$ in simulation in a $4\times 20$ lattice, with the initial excitation source being $s=e^{-i\Delta kz}$ with $\Delta k=-2g$ applied in the 11th mode of waveguide 1 denoted by the arrow, to excite the exact edge state with blue dot in (a). Depicted is the field profile at $z=40g^{-1}$. (d) The average intensity spectra $I_{n,l}$ for the probe field at each waveguide.}
\label{fig2}
\end{figure*}
As the first demonstration, we begin our numerical analysis by considering the four-waveguide system with independent pump beams. In other words, we ignore the crosstalk between pump fields for the moment so each pump beam only affects the probe field in each waveguide, i.e., $\kappa_l(z) = \kappa_l(0) \equiv \kappa$. In this case, the system is exactly described by the Hamiltonian (\ref{seq4}). The topological property of this synthetic system manifest itself in the existence of topologically chiral edge states crossing the bandgap. By assuming infinite modes along the frequency axis, we calculate the projected band structurewith a Hamiltonian versus $k_f$, which is reciprocal to the frequency dimension\cite{yuan2021tutorial}:
\begin{equation}
\label{partialFT}
  H_{k_f}=\sum_{\substack{l}} \left[2\kappa a_{k_f,l}^\dag a_{k_f,l} \cos(k_f\Omega_f-\phi_l)+g\left(a_{k_f,l}^\dag a_{k_f,l+1}+a_{k_f,l+1}^\dag a_{k_f,l}\right)\right].
\end{equation}
The projected band structure is shown in Fig. 2(a). 
Since we have 4 waveguides that support a synthetic lattice with only 4 rows[see Fig. \ref{fig1}(b)], 4 separate bands can be seen in Fig. \ref{fig2}(a). Yet, the chiral edge states still exist, of which the field intensity of the states decay exponentially into the bulk [Fig. \ref{fig2}(b)]. To better study the characteristic of the edge states, we simulate the transport of the wave function of light in a $4\times 20$ synthetic lattice, i.e., only 20 frequency modes are considered in simulations. Artificial boundaries at the frequency dimension are considered, which is possible to be achieved by engineering the dispersion of the waveguide\cite{yuan2017synthetic,shan2020one}. We write the wave function of light as\cite{yuan2021tutorial}
\begin{equation}
 |\Psi(z)\rangle=\sum_{\substack{n,l}}v_{n,l} a_{n,l}^\dag |0\rangle,\label{wf}
\end{equation}
in which $v_{n,l}$ is the probability amplitude of the photon state at site $(n,l)$.  By substituting Eq. (\ref{wf}) into the Schr\"odinger-like equation $id|\Psi(z)\rangle/dz=H|\Psi(z)\rangle$, we find the governing working equation is
\begin{equation}
\label{ode}
 \dfrac{d}{d z} v_{n,l}=-i\kappa\left(v_{n+1,l}e^{-i\phi_l}+v_{n-1,l}e^{i\phi_l}\right)-ig\left(v_{n,l+1}+v_{n,l-1}\right)+s\delta_{n,11}\delta_{l,1}.
\end{equation}
Note that to excite the system, a source excitation $s=e^{-i\Delta kz}$ has been adopted which is applied to excite the $11$th frequency mode in the first waveguide. Here, $\Delta k$ is the wavevector mismatching with respect to that of the considered mode\cite{yuan2017synthetic}, and such excitation can be achieved by using the travelling excitation field propagating inside an additional waveguide weakly coupled to the first waveguide of the system\cite{stutzer2018photonic}. In Fig. \ref{fig2}(c), we show the calculated distribution of the intensity $|v_{n,l}|^2$ in the synthetic lattice at the propagation distance $z=40g^{-1}$.
One can see a one-way propagation of the edge state in the synthetic space, i.e., the edge state propagates unidirectionally along the boundaries of the synthetic lattice. 
It is instructive to investigate the average intensity spectra, $I_{n,l} \equiv \int |v_n,l|^2 dz$. The values of $|v_{n,l}|^2$ are related to the number of photons at each  lattice site $(n, l)$, and therefore $I_{n,l}$ gives the average intensity spectra for the probe field propagate through each entire waveguide, which is plotted in Fig. \ref{fig2}(d). Our results here show that the energy  of the probe is confined on the boundary of the synthetic lattice and cannot diffuse into the bulk in the pumped waveguide system.

\subsection{Coupled-mode analysis on the crosstalk of pump fields}
\begin{figure*}
\centering\includegraphics[width=8.5cm]{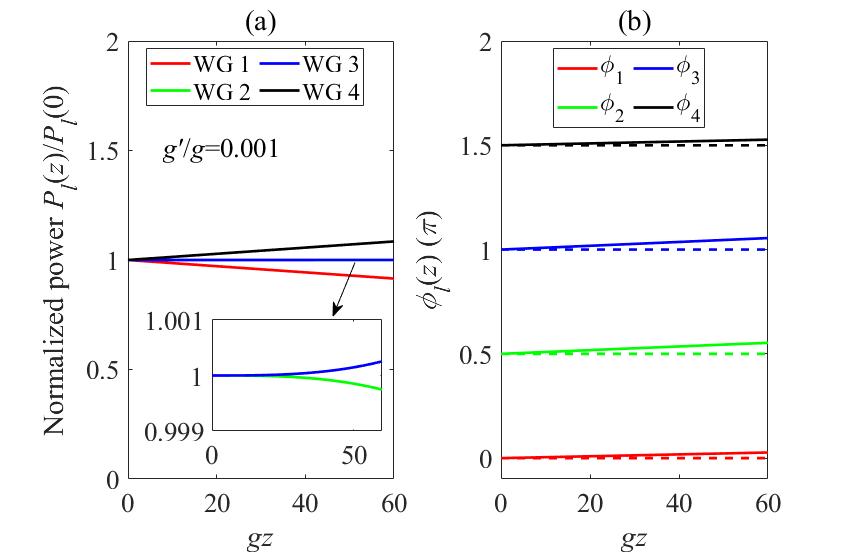}
\centering\includegraphics[width=8.5cm]{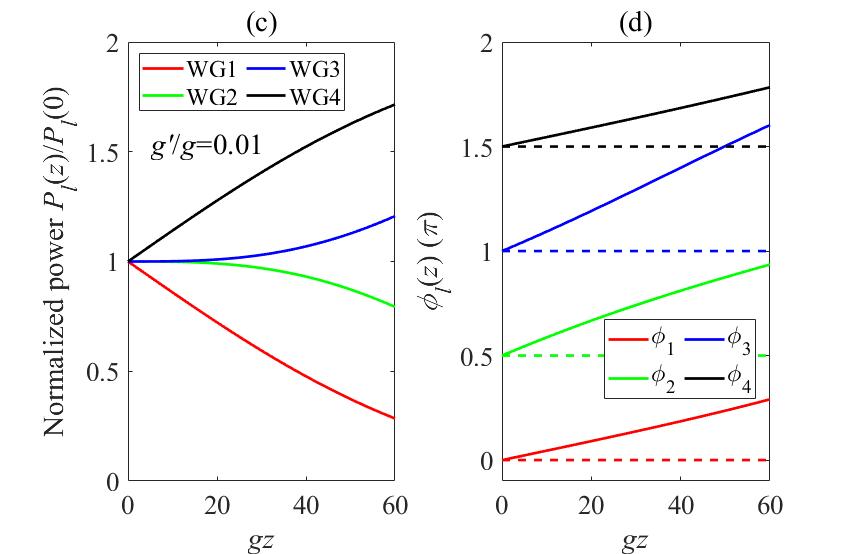}
\caption{(Color online) Simulation results of powers $P_l(z)$ and the resulting phases $\phi_l(z)=$2Arg$[u_l^p(z)]$ of pumps in each waveguide (WG). (a) and (b) $g'/g=0.001$; (c) and (d) $g'/g=0.01$.}
\label{fig3}
\end{figure*}

To explore the role of the crosstalk between pumps, we now perform the analysis based on the coupled mode theory\cite{yariv1973coupled,haus1991coupled,huang1994coupled} to study the interactions of pump fields in different waveguides. In particular, we numerically solve Eq. (\ref{peq1}) with initial conditions $[u^p_{1}(0),u^p_{2}(0),u^p_{3}(0),u^p_{4}(0)]=u_0[1, $exp$(i\pi/4), $exp$(i\pi/2), $exp$(3i\pi/4)]$. This initial condition for pumps indicates that four waveguides are pumped by optical fields at the same intensity but different phases, which is exactly what we used in the previous subsection. However, here, we consider changes of pumps propagating through waveguides due to the crosstalk.
We take two examples under different coupling coefficients for the pumps: $g'/g=0.001$ and $g'/g=0.01$. 
In Fig. \ref{fig3}, We plot the simulation results of powers $P_l(z)$ and the resulting phases $\phi_l(z)=$2Arg$[u_l^p(z)]$ of pumps in each waveguide versus the propagating distance $z$.

Fig. \ref{fig3}(a) clearly shows that powers $P_l$ become position dependent for $g'/g=0.001$. The pump powers in waveguide 1 and 2 decrease from their initial values $P_{1,2}(0)=|u_0|^2$ in the simulation range of $gz\in[0,60]$. At the same time, the pump powers in waveguide 3 and 4 increase. Notice that as compared with the changes of the pump powers in waveguides 1 and 4, those in waveguides 2 and 3 change more slowly, which can been seen in the inset of Fig. \ref{fig3}(a). The reason is that the initial phases play a significant role in the redistribution of energy in the waveguide array through the interference among the amplitudes of the pumps. Compared with the case of the simulation with initially with no phase difference, the pump field in waveguide 1(2) changes equivalently to that in waveguide 4(3). However, for the case with initial phase difference, such degeneracy due to the symmetry is lifted as shown in Fig. \ref{fig3}(a).
As the coupling coefficient increases, such as $g'/g=0.01$, the impact of the crosstalk gets larger, which is illustrated in Fig. \ref{fig3}(c). 

The crosstalk of the pump fields not only affects the power redistributions, but more importantly, it brings disorder into the phase distribution of the pump field inside each waveguide, as plotted in Figs. \ref{fig3}(b) and \ref{fig3}(d).
Comparing the results with different amplitudes of $g'$ in Fig. \ref{fig3}(b) and \ref{fig3}(d), one can see that for weak coupling case $g'/g=0.001$, the phases for pump fields in each waveguide experience a small linear change in the propagation range of $gz=60$. Moreover, phases in waveguides 1 and 4 change more slowly than those in waveguides 2 and 3. As the coupling coefficient of the pump fields increases, all phases have been changed dramatically within the same propagation distance. Therefore, the coupled-mode analysis explicitly shows that the evolution of the powers and phases of the pump fields depends on both the coupling coefficient $g'$ and the propagation distance $z$, which shall affect the dynamics of edge states for the probe field.  

\subsection{All-optical control of edge states}
 \begin{figure*}
\centering\includegraphics[width=5.0cm]{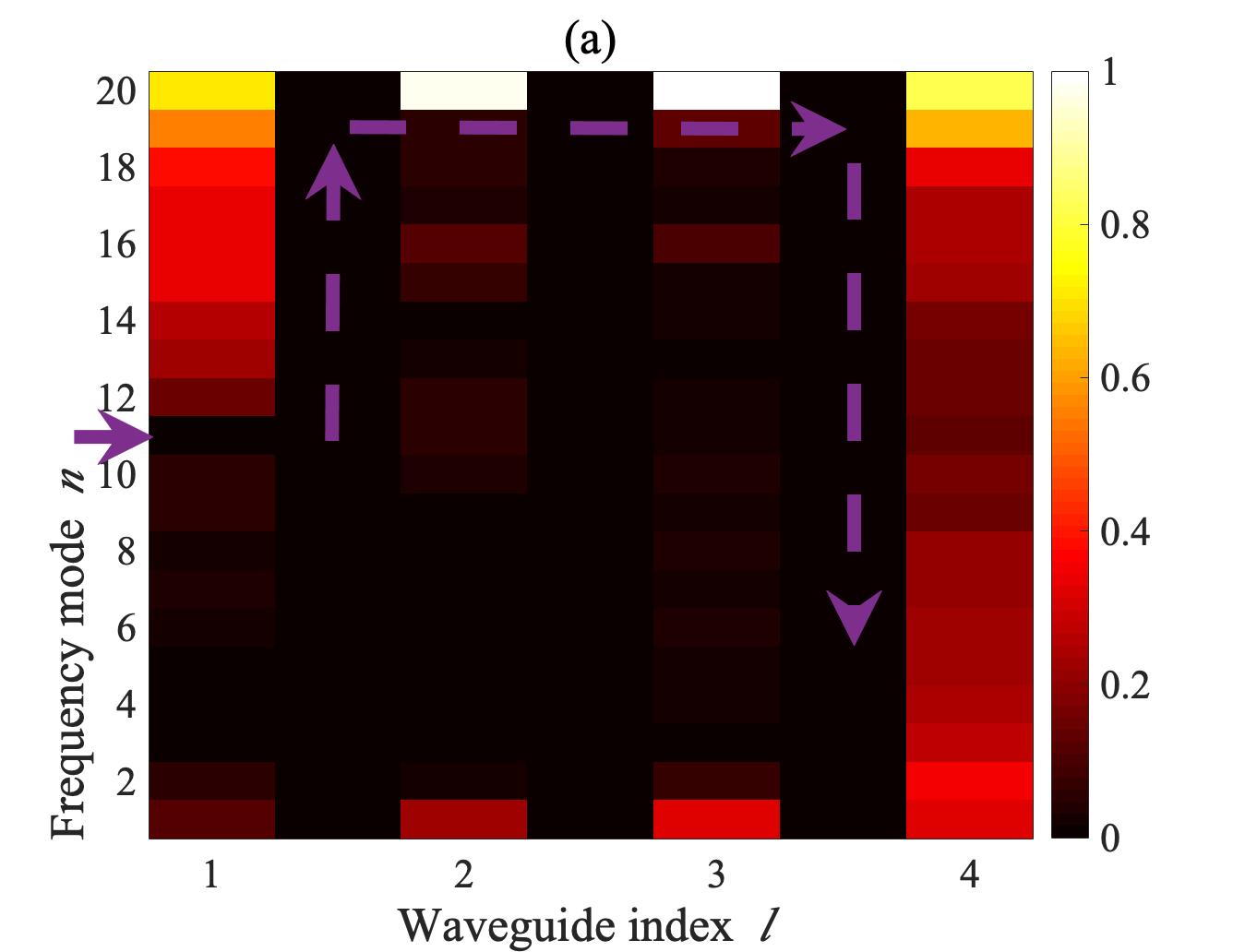}
\centering\includegraphics[width=5.0cm]{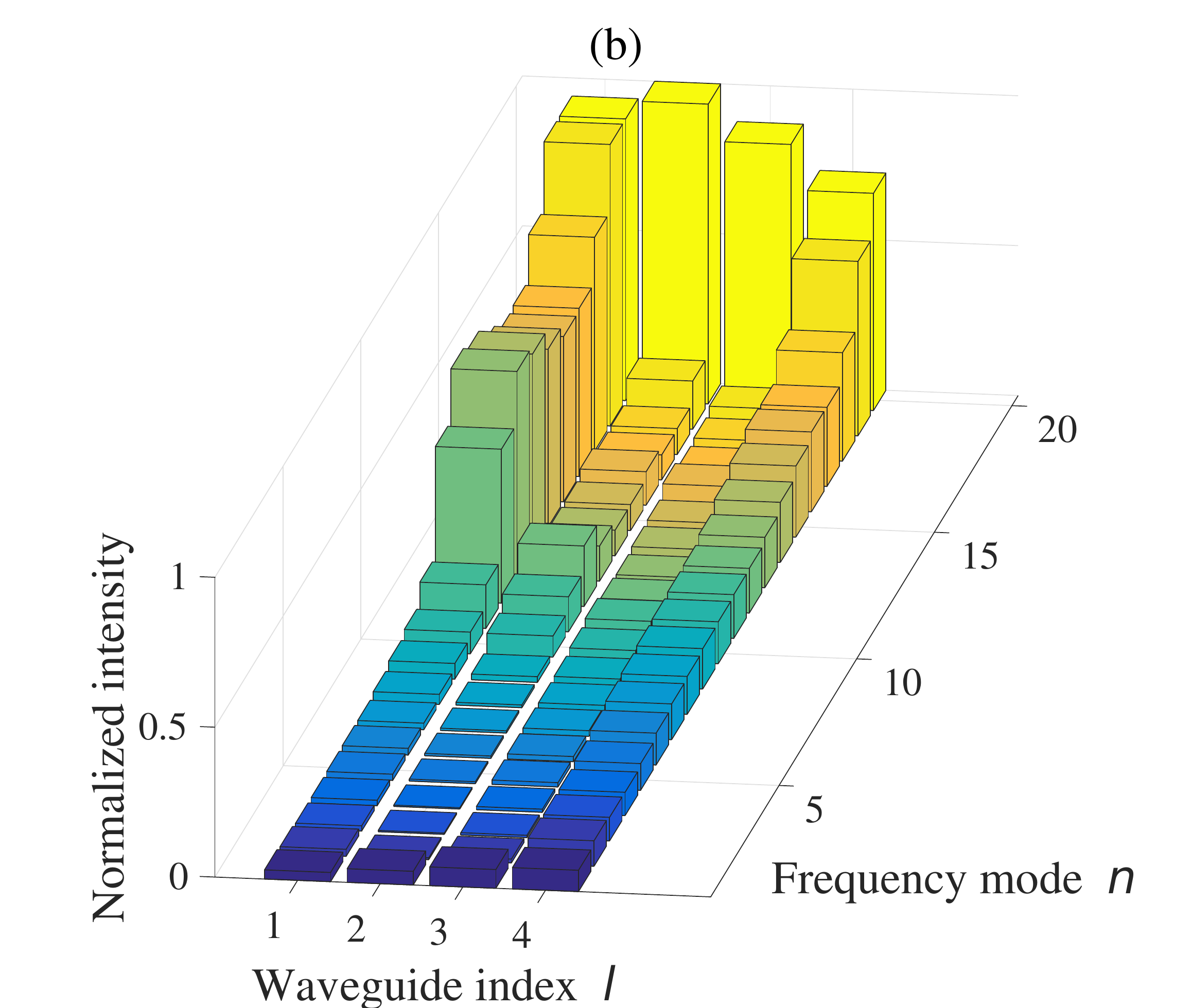}
\centering\includegraphics[width=5.0cm]{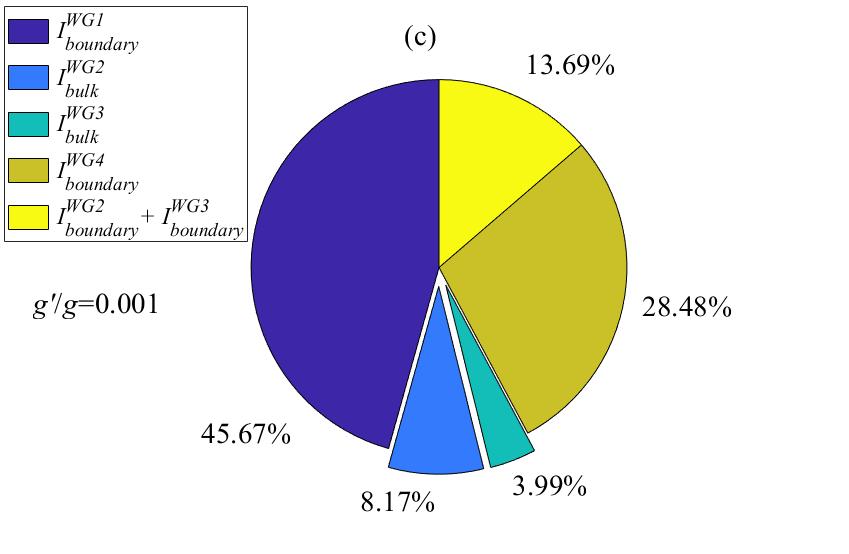}
\centering\includegraphics[width=5.0cm]{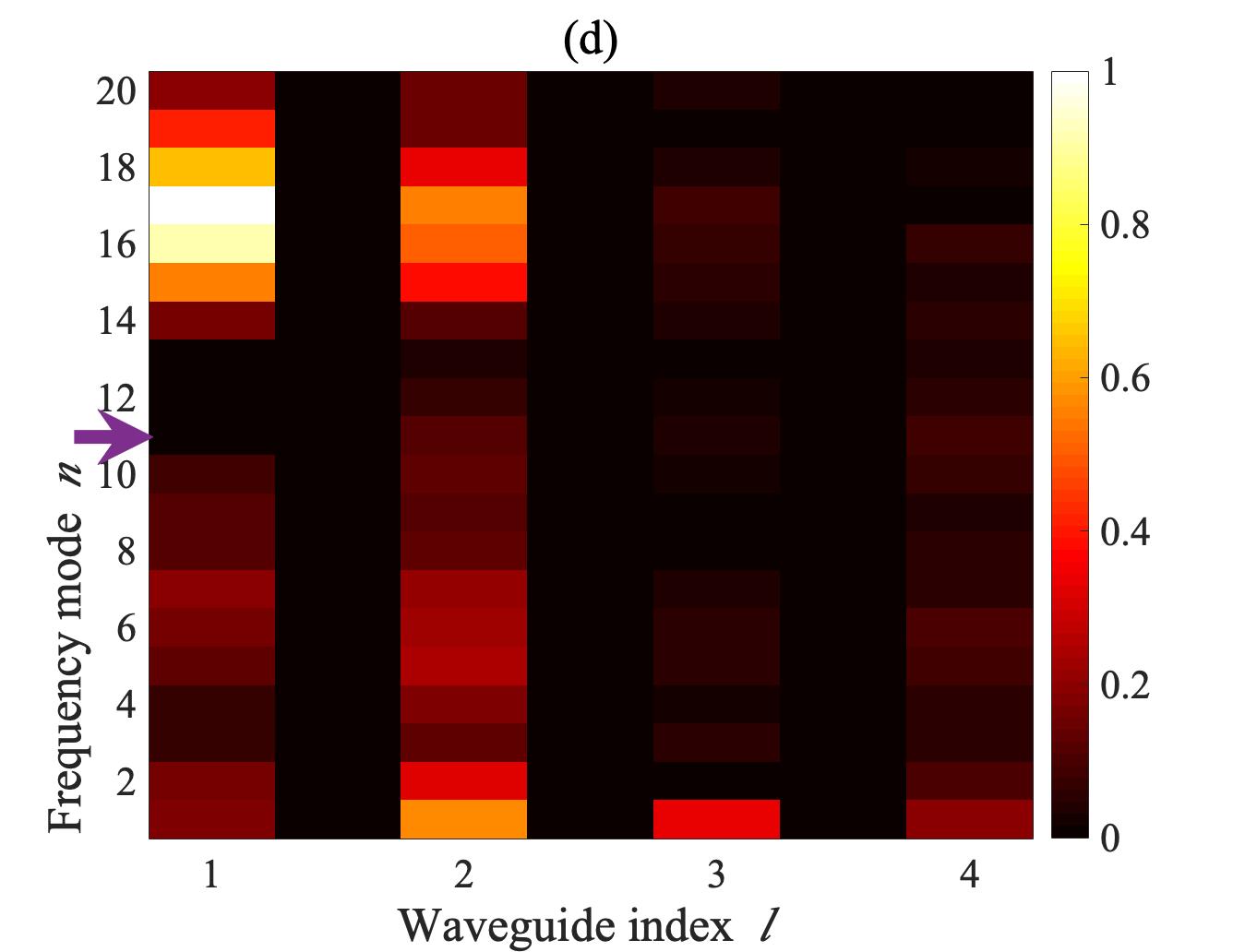}
\centering\includegraphics[width=5.0cm]{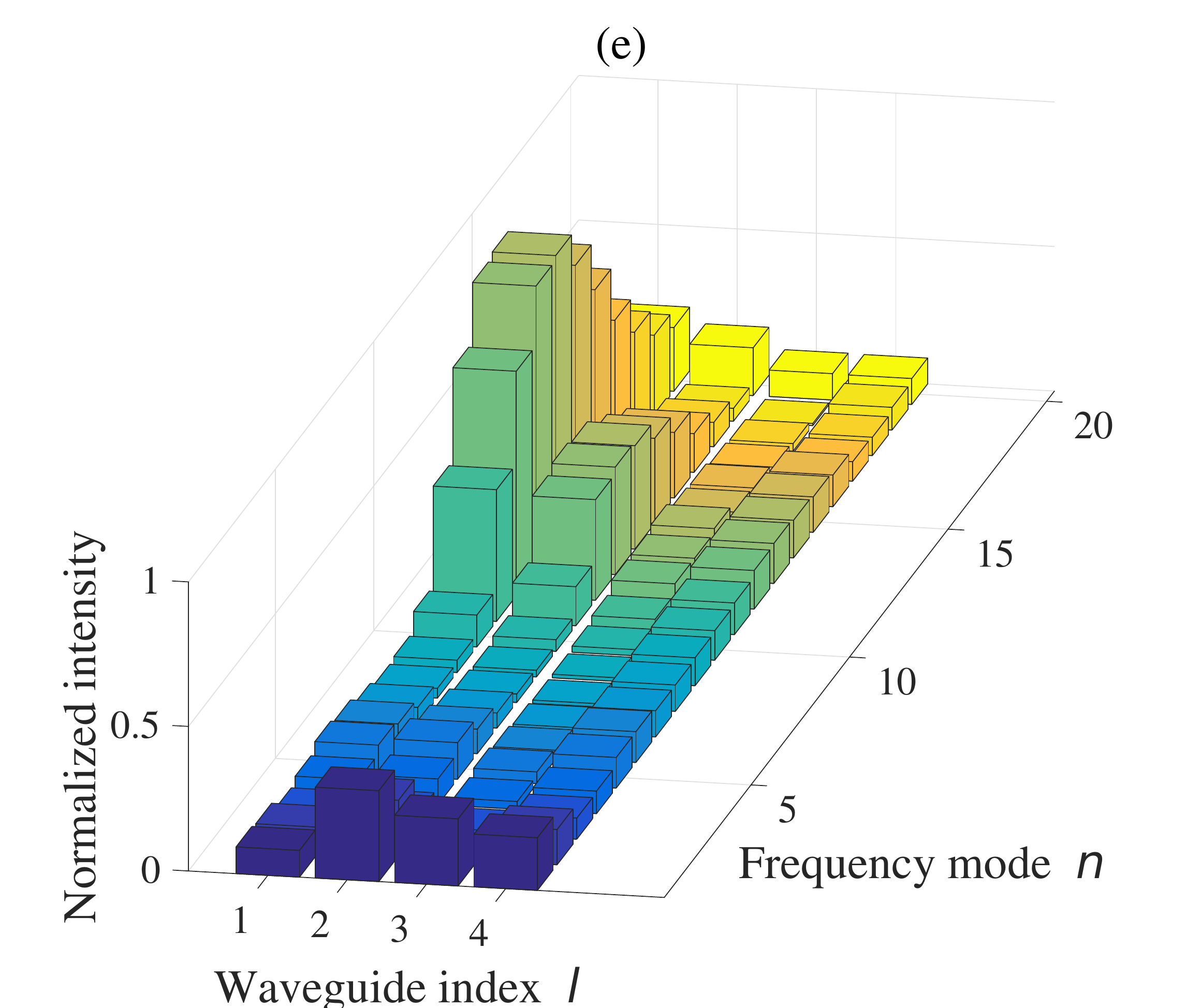}
\centering\includegraphics[width=5.0cm]{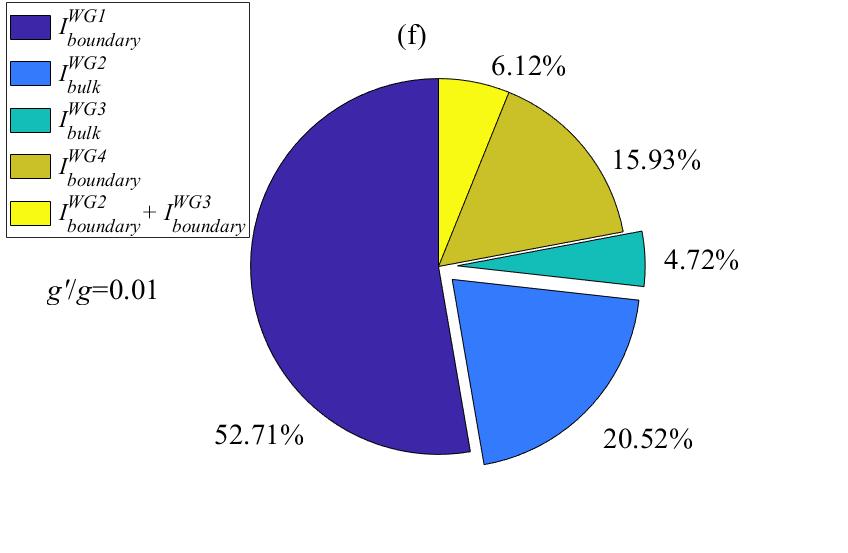}
\centering\includegraphics[width=6.0cm]{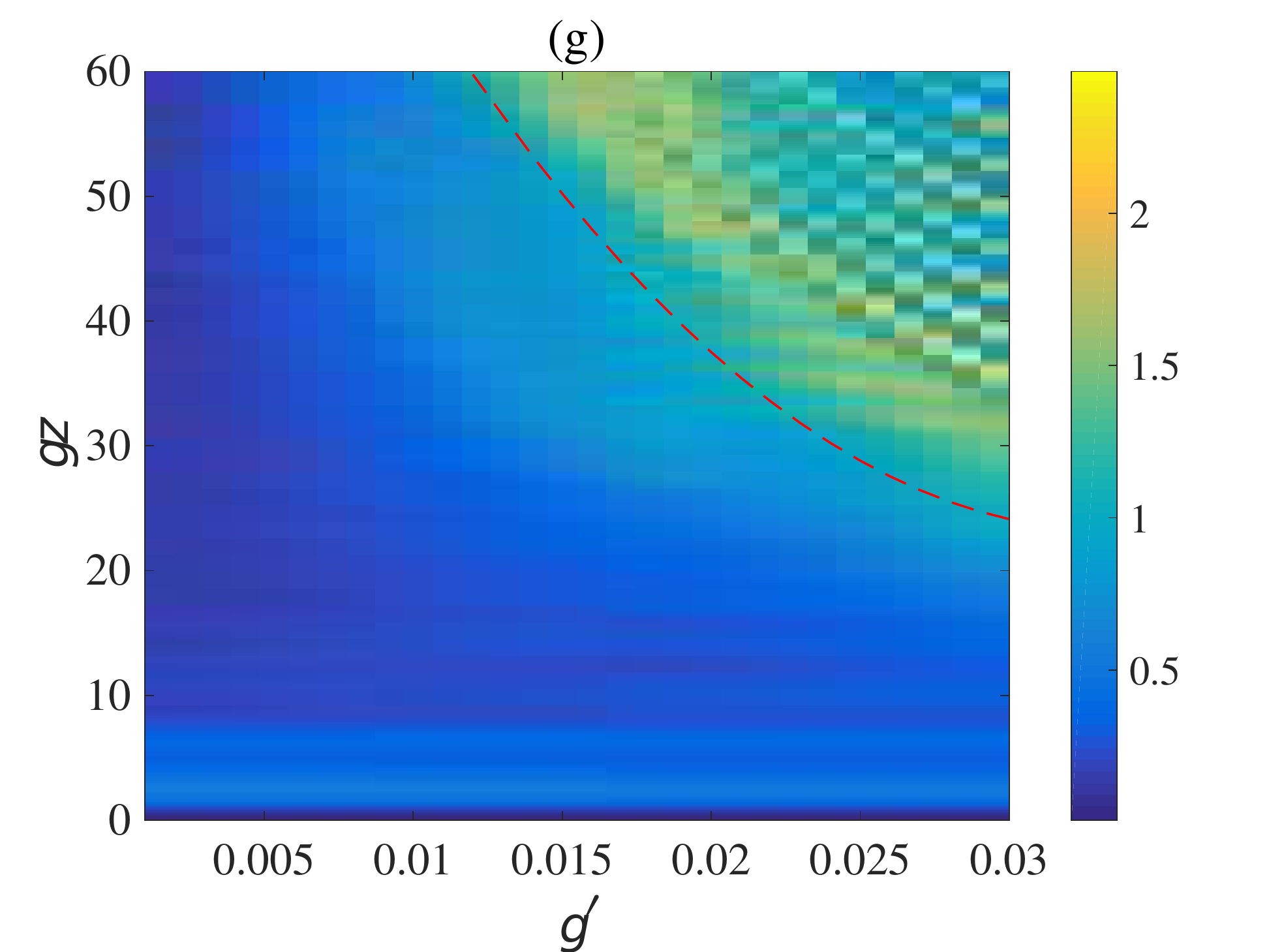}
\centering\includegraphics[width=6.0cm]{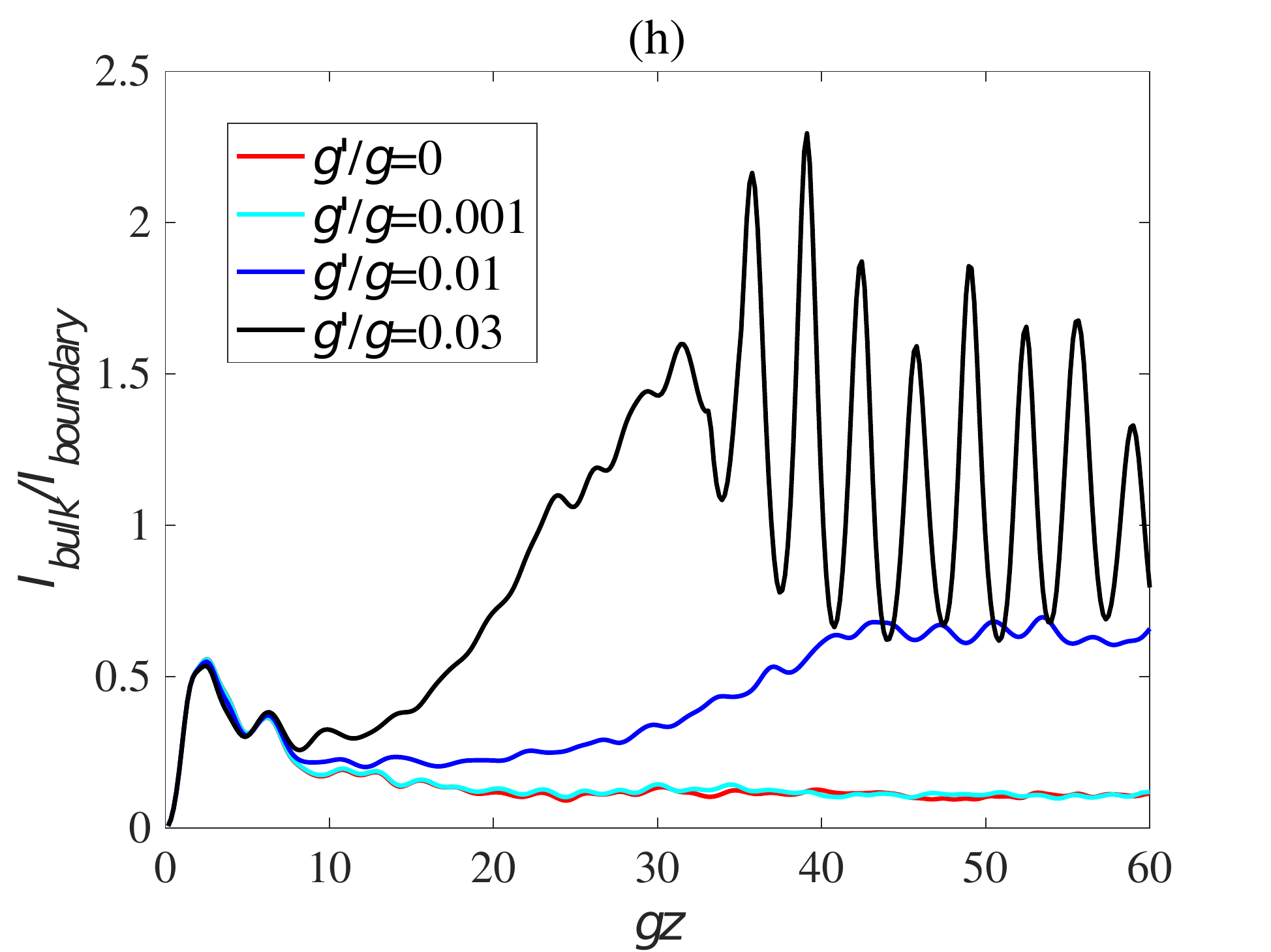}
\caption{(Color online)  Simulation results in two cases with $g'/g=0.001$ (a)-(c), and with $g'/g=0.01$ (d)-(f), under the same excitation as discussed in Fig. \ref{fig2}. (a) and (d) Distributions of $|v_{n,l}|^2$ for the probe field at $z=40g^{-1}$. (b) and (e) Average intensity spectra $I_{n,l}$ for the probe field at each waveguide. (c) and (f) Distributions of the energy in different bulk and boundary parts. (g) Phase diagram of $I_{bulk}/I_{boundary}$ as a function of the propagation distance and the coupling coefficient for the pump fields. Red dashed line denotes $I_{bulk}/I_{boundary}\sim 1$. (h) Plots of $I_{bulk}/I_{boundary}$ as a function of the propagation distance for cases with $g'/g=0, 0.001, 0.01, 0.03$.}
\label{fig4}
\end{figure*}
We are now in a position to take the coupled-mode results for the pump fields into account and explore its effect on the probe propagation. 
The fundamental difference between the Hamiltonian with the coupled pump fields in Eq. (\ref{seq6}) and the Hamiltonian  (\ref{seq4}) that we studied before is that hopping coefficients for the probe field $\kappa_l(z)$ and phases $\phi_l(z)$ in the $l$-th waveguide is no longer  uniform and identical, which are dependent on the pump field $u_l^p(z)$ in the arrays.

Firstly, we study the small coupling case, i.e., $g'/g=0.001$. 
We perform simulations with the Hamiltonian (\ref{seq6}) using same parameters as those for Fig. \ref{fig2} (c) and \ref{fig2} (d), and plot results in Fig. \ref{fig4} (a)-\ref{fig4} (c).
 Fig. \ref{fig4} (a) plots the distribution of the intensity $|v_{n,l}|^2$ in the synthetic lattice at the propagation distance $z=40g^{-1}$. Due to the crosstalk between the pump fields in different waveguides, it shows small deviation from the results in Fig. \ref{fig2} (c) where the crosstalk is ignored. This small difference in two cases can also be seen from the average intensity spectra $I_{n,l}$ plotted in Fig. \ref{fig4}(b), which shows that the most part of the energy of the probe field is confined to the boundary of the synthetic lattice.
To better explore the energy distribution, the boundary region and the bulk region of the synthetic lattice structure are separated, and the average intensity distributed in the boundary and the bulk sites are defined by $I_{boundary}$ and $I_{bulk}$, respectively. Specifically, the boundary region corresponds to all the frequency modes in waveguides 1 and 4, as well as the highest and lowest frequency modes in waveguides 2 and 3, while the rest modes in waveguides 2 and 3 contribute to the bulk region. Hence, following this specific separation rule we can define 
$I_{boundary}=I_{boundary}^{WG1}+I_{boundary}^{WG2}+I_{boundary}^{WG3}+I_{boundary}^{WG4}$, and $I_{bulk}=I_{bulk}^{WG2}+I_{bulk}^{WG3}$, where the distribution of the average intensity inside each part is depicted in Fig. \ref{fig4}(c). One can see that the boundary part $I_{boundary}$ contributes about $88\%$ of the total energy in which the frequency modes in waveguide 1 ($I_{boundary}^{WG1}$) account for more than half of the energy of the edge state, whereas the bulk part $I_{bulk}$ occupies $\sim 12\%$ of the total energy. We thus find that the probe field exhibits its edge state under the influence of a small crosstalk effect ($g'/g=0.001$) from the pump fields.

Secondly, we consider the case with the significant coupling coefficient for the pump fields, e.g., $g'/g=0.01$.  
Fig. \ref{fig4}(d) shows the intensity $|v_{n,l}|^2$ of the probe field at the propagation distance $z=40g^{-1}$. One finds that, at this distance, the probe field propagates into the bulk region dramatically. 
More specifically, the energy distribution in the synthetic lattice tends to confine to all frequency modes in waveguide 1 and waveguide 2. Fig. \ref{fig4}(e) plots the average intensity spectra $I_{n,l}$, which confirms this tendency.
In Fig. \ref{fig4}(f), we show the distribution of average intensity inside each part we defined in the previous paragraph, where we find that the coupling coefficient for the pump fields clearly redistributes energy in the boundary region and the bulk region. By comparing the case with $g'/g=0.001$, we find that more energy has been diverted from boundary parts to bulk parts, resulting $\sim25\%$ in the bulk region.
The important result in the case with $g'/g=0.01$, as shown in Fig. \ref{fig4}(e), however, is that the probe field gets the unidirectional frequency conversion and obtains the peak power in the vicinity of the frequency mode $n=15$ in waveguide 1, while the energy of the probe penetrates into the bulk near the 15th mode in the waveguide 2. Such phenomenon comes from the competition between the topological effect and the disorder in the modulation strengths and the phases for the probe due to the crosstalk from the pump fields.

To illustrate the influence of the crosstalk from the pump fields in this all-optical configuration in more details, we present the results of the numerically calculated ratio of the average intensity of the frequency modes in the bulk region to those in the boundary region, i.e., $I_{bulk}/I_{boundary}$, as a function of the propagation distance and the coupling coefficient for the pump fields, $g'$, as shown in Fig. \ref{fig4}(g). It serves as a quantitative indicator to investigate the edge-mode dynamics. More specifically, when the value of $I_{bulk}/I_{boundary}$ is smaller(larger) than 1, it represents that the boundary(bulk) modes dominates at the parameter space $(z, g’)$. 
This phase diagram in Fig. \ref{fig4}(g) shows that there exhibits edge state for small $g'$ or short propagation distance. In such parameter space, $I_{boundary}$ changes continuously.
However, in the region that $I_{bulk}/I_{boundary}>1$, this ratio undergoes oscillations in the parameter space with large $g'$ and long propagation distance. We examine such behavior of the ratio
$I_{bulk}/I_{boundary}$ with different choices of $g'$ in Fig. \ref{fig4}(h). One can see that,
for $g'=0$, the propagating modes of the probe field on the boundary of the synthetic lattice overwhelmingly dominates. In the case with $g'/g=0.001$, similar behavior can be seen with small deviation. As the value of $g'/g$ increases, more energy is transferred from the boundary region to the bulk region during the probe propagates. While it is still the edge-mode dominates for the case with $g'/g=0.01$ in the propagation range $gz=60$, in the case with $g'/g=0.03$ the ratio $I_{bulk}/I_{boundary}$ transit quickly to the bulk-mode dominated region and oscillates with $z$ from the reflection at boundaries in the synthetic lattice. In this case, due to the relatively large crosstalk between the pump fields in different waveguide during the propagation, the chiral edge state cannot be maintained, making the breakdown of the quantum Hall phase.

\section{Discussion and conclusion}
Experimentally, the synthetic frequency axis of light via cross-phase modulation has been observed in an optical fiber system\cite{bersch2009experimental,bersch2011spectral}. The demonstration of our theoretical proposal consisting of a frequency dimension and a spatial dimension requires an array of coupled waveguides, which can be fabricated using the optical induction technique\cite{efremidis2002discrete,fleischer2003observation} or the femtosecond laser direct writing technique\cite{szameit2005discrete,szameit2006two}.
To ignore the effective electric filed in the synthetic two-dimensional lattice, it needs the condition that $\gamma P_0\gg n\Delta\Omega$, which can matched by either choosing materials with large optical nonlinearity or high pump power, or considering small walk-off parameter or modulation frequency.   
On the other hand, in order to control the crosstalk from the pump optical fields,
one can vary the wavelength of the pump fields so that the coupling coefficient of the pump fields in the waveguide array changes accordingly\cite{iwanow2005discrete,solntsev2012spontaneous,yu2012spatiospectral,kruse2013spatio}. Hence, the desired wavelength window allows one to tune the ratio of coupling coefficients $g'/g$. Moreover, there are several alternative experimental approaches which have been demonstrated to be able to control the crosstalk in waveguide systems, such as examples of using transformation optics\cite{gabrielli2012chip}, anisotropic metamaterials\cite{jahani2018controlling}, and methods inspired from the atomic physics\cite{song2015high,mrejen2015adiabatic}.
Therefore, it is possible to construct the two-dimensional synthetic lattice in a waveguide array and control the crosstalk in such a platform under the current photonic technology, making our proposal feasible to be experimentally realized. 

In summary, we study the light transport in a quantum Hall system in the synthetic space, consisting of the probe field propagating in a one-dimensional four-waveguide array driven by pump fields at the designed phase distribution. The interaction between the pump and probe field via cross-phase modulations inside each waveguide induces a synthetic frequency dimension, and moreover, the phase distribution of the pump fields provides an alternative all-optical path towards the realization of artificial gauge potential and hence effective magnetic field for photons in the pumped waveguide array. 
We explore the dynamics of the topological chiral edge states and also the edge-mode dynamics affected by the crosstalk of pump lasers, where the parameter space has been investigated to seek potential controllability of the probe flow in the synthetic space.
Besides the four-waveguide system, our study can also be generalized to higher-dimensional waveguide arrays.
The all-optically control of chiral transport of light on the boundary in a synthetic space can be of great importance in nonlinear frequency generation and signal multiplexing for communication\cite{chen2021highlighting}.

\begin{acknowledgments}
The research is supported by National Natural Science Foundation of China (12122407, 11974245, and 12104296), National Key R\&D Program of China (2017YFA0303701), Shanghai Municipal Science and Technology Major Project (2019SHZDZX01), and Natural Science Foundation of Shanghai (19ZR1475700). L.Y. acknowledges the support from the Program for Professor of Special Appointment (Eastern Scholar) at Shanghai Institutions of Higher Learning. X.C. also acknowledges the support from Shandong Quancheng Scholarship (00242019024).
\end{acknowledgments}

\bibliography{myref.bib}

\end{document}